\documentclass[epj,referee]{svjour}
\usepackage[dvips]{graphicx}
\newcommand{\eqref}[1] {(\ref{#1})}
\begin{document}
\title{
Impact of two mass-scale oscillations on the analysis
of atmospheric and reactor neutrino data}
\author{M.~C.~Gonzalez-Garcia \inst{1,2,3} and M. Maltoni \inst{2}}
\institute{Theory Division, CERN, CH-1211 Geneva 23, Switzerland
\and Instituto de F\'{\i}sica Corpuscular,
  Universitat de  Val\`encia -- C.S.I.C.\\
  Edificio Institutos de Paterna, Apt.~22085, E--46071 Valencia, Spain
\and C.N. Yang Institute for Theoretical Physics\\
  State University of New York at Stony Brook}

\abstract
{We study the stability of the results of 3-$\nu$ oscillation analysis of
atmospheric and reactor neutrino data under departures of the
one--dominant mass scale approximation. In order to do so 
we perform  the analysis of atmospheric and reactor neutrino data 
in terms of three--neutrino oscillations where the effect of both
mass differences is explicitly considered. We study the allowed
parameter space resulting from this analysis as a function of the
{\it mass splitting hierarchy parameter} $\alpha=\Delta m^2/
\Delta M^2$ which parametrizes the departure from the one--dominant
mass scale approximation. We consider schemes with both direct and
inverted mass ordering. Our results show that in the analysis
of atmospheric data  the derived range of the largest mass splitting, 
$\Delta M^2$, is stable while the allowed
ranges of mixing angles $\sin^2\theta_{23}$ and 
$\sin^2\theta_{13}$ are wider than those obtained in
the one--dominant mass scale approximation. Inclusion of the CHOOZ
reactor data in the analysis results into the reduction of the 
parameter space in particular for the mixing angles. 
As a 
consequence the final allowed ranges of parameters from the combined 
analysis are only slightly broader than when obtained in the 
one--dominant mass scale approximation.}
\maketitle

\section{Introduction}

Super--Kamiokande (SK) high statistics data~\cite{skatmlast} indicate
that the observed deficit in the $\mu$-like atmospheric events is due
to the neutrinos arriving in the detector at large zenith angles,
strongly suggestive of the $\nu_\mu$ oscillation hypothesis.
Similarly, the latest SNO results~\cite{sno,snonc} in combination  with the 
Super--Kamiokande data on the zenith angle dependence and recoil energy
spectrum of solar neutrinos~\cite{sksollast} and the 
Homestake~\cite{chlorine}, SAGE~\cite{sage}, and
GALLEX+GNO~\cite{gallex,gno} experiments, have put on a firm
observational basis the long--standing problem of solar
neutrinos~\cite{snp}, strongly indicating the need for $\nu_e$
conversions.

Altogether, the solar and atmospheric neutrino anomalies constitute
the only solid present--day evidence for physics beyond the Standard
Model \cite{review}. It is clear that the minimum joint description of both
anomalies requires neutrino conversions among all three known
neutrinos. In the simplest case of oscillations the latter are
determined by the structure of the lepton mixing matrix~\cite{MNS},
which, in addition to the Dirac-type phase analogous to that of the
quark sector, contains two physical phases associated to the Majorana
character of neutrinos, which however are not relevant for neutrino
oscillation \cite{majophases} 
and will be set to zero in what follows. In this case the
mixing matrix $U$ can be conveniently chosen in the form~\cite{PDG}
\begin{equation}
\left(
    \begin{array}{ccc}
        c_{13} c_{12}
        & s_{12} c_{13}
        & s_{13} \,{e}^{-i\delta}\\
        -s_{12} c_{23} - s_{23} s_{13} c_{12} \,{e}^{i\delta}
        & c_{23} c_{12} - s_{23} s_{13} s_{12}\, {e}^{i\delta}
        & s_{23} c_{13} \\
        s_{23} s_{12} - s_{13} c_{23} c_{12} \,{e}^{i\delta}
        & -s_{23} c_{12} - s_{13} s_{12} c_{23} \, {e}^{i\delta}
        & c_{23} c_{13}
    \end{array} \right) \,,
    \label{eq:matrix}
\end{equation}

where $c_{ij} \equiv \cos\theta_{ij}$ and $s_{ij} \equiv
\sin\theta_{ij}$. Thus the parameter set relevant for the joint study
of solar and atmospheric conversions becomes six-dimensional: two mass
differences, three mixing angles and one CP phase.

Results from the analysis of solar and atmospheric data in the
framework of two-neutrino
oscillation~\cite{2solour,2solother,2atm,3atmfogli} imply that the
required mass differences satisfy
\begin{equation}
    \Delta m^2_\odot
    \ll \Delta m^2_{\rm atm}.
    \label{eq:deltahier}
\end{equation}
For sufficiently small $\Delta m^2_\odot$ the three-neutrino
oscillation analysis of the atmospheric neutrino data can be performed
in {\it the one mass scale dominance approximation} neglecting the
effect of $\Delta m^2_\odot$. In this approximation it follows that
the atmospheric data analysis restricts three of the oscillation
parameters, namely, $\Delta m^2_{31} = \Delta m^2_{32}$, $\theta_{23}$
and $\theta_{13}$. This is the approximation used in 
Ref.~\cite{3atmfogli,3ours,3atm}. Conversely for the
solar neutrino analysis the effect of oscillations with $\Delta
m^2_{\rm atm}$ can be taken to be averaged and solar data constrains
$\Delta m^2_{21}, \theta_{12}$ and $\theta_{13}$~\cite{3ours,3sol}. In
this approximation the reactor neutrino data from CHOOZ provides
information on the atmospheric mass difference and the mixing angle
$\theta_{13}$, and the CP phase $\delta$ becomes unobservable.

However the assumption of one mass scale dominance may not be a good
approximation neither for reactor nor for atmospheric data, in
particular for $\Delta m^2_\odot$ in its upper allowed values. Effects
of the departure of the one mass scale dominance approximation in the
analysis of the CHOOZ reactor data~\cite{CHOOZ} has been included in
Ref.~\cite{3ours,choozpetcov,irina}. For atmospheric neutrinos in
Refs.~\cite{strumia,3atmfogli,antonio} it was shown that oscillations 
with two
mass scales of the order of $10^{-3}$ could give a good description of
the existing data for some specific values of the parameters.  
Some analytical approximate expressions for the effects of keeping 
both mass scales in the description of atmospheric neutrinos are presented in
Refs.~\cite{smirnov1,smirnov2,nohierothers}.
Furthermore Refs.~\cite{smirnov1,smirnov2} describe how   
the presence of the second mass scale can lead to an 
increase in the number of sub-GeV electron events which seems to
improve the description of the observed distribution.

To further explore this possibility and to verify the consistency
of the one--dominant mass scale approximation we present in this work 
the result of the analysis of the atmospheric and reactor neutrino 
data in terms of three-neutrino oscillations where the effect of both
mass differences is explicitly considered and  we compare our results
with those obtained under the assumption of one-dominant scale.  
Our aim is to study how/whether the allowed parameter space is modified as 
a function of the ratio between the two mass scales.  
Our study allow us to establish the 
stability of the derived ranges of parameters for the large mass
scale and mixings  $\theta_{23}$ and $\theta_{13}$
{\sl independently} of the exact 
value of the solar small scale and mixing $\theta_{12}$ for which
we only chose it to be within the favoured LMA region. 
Our results show that the allowed ranges of parameters from 
the combined atmospheric plus reactor data analysis are only 
slightly broader than when obtained in the one--dominant mass scale 
approximation. Thus our main conclusion is that the 
approximation is self-consistent. To establish the relevance of 
each data sample on this conclusion we also present the partial 
results of the analysis including only the atmospheric data or 
the reactor data.

The outline of this paper is as follows. In Sec.~\ref{sec:parameters}
we describe our notation for the parameters relevant for atmospheric
and reactor neutrino oscillations with two mass scales and discuss
the results for the relevant probabilities. In
Sec.~\ref{sec:analatmos} and~\ref{sec:analchooz} we show our results
for the analysis of atmospheric neutrino and reactor data
respectively. For atmospheric neutrinos we include in our analysis all the
contained events from the latest 1489 SK data
set~\cite{skatmlast}, as well as the upward-going neutrino-induced
muon fluxes from both SK and the MACRO detector~\cite{MACRO}.
The results for the combined analysis are described in
Sec.~\ref{sec:combined}.  Finally in Sec.~\ref{sec:sum} we summarize
the work and present our conclusions.

\section{Three Neutrino Oscillations with Two Mass Scales}
\label{sec:parameters}

In this section we review the theoretical calculation of the
conversion probabilities for atmospheric and reactor neutrinos in the
framework of three--neutrino mixing, in order to set our notation and
to clarify the approximations used in the evaluation of such
probabilities.

In general, the determination of the oscillation probabilities 
for atmospheric neutrinos require the solution of the Schr\"odinger
evolution equation of the neutrino system in the Earth--matter
background. For a three-flavour scenario, this equation reads
\begin{equation}
    i \frac{d\vec{\nu}}{dt} = {\bf H} \, \vec{\nu}, \qquad
    {\bf H} = {\bf U} \cdot {\bf H}_0^d \cdot {\bf U}^\dagger
    + {\bf V} \,,
    \label{eq:evol.1}
\end{equation}
where ${\bf U}$ is the unitary matrix connecting the flavour basis and
the mass basis in vacuum and which can be parametrized as in
Eq.~\eqref{eq:matrix}. On the other hand ${\bf H}_0^d$ and {\bf V} are
given as
\begin{equation}
    {\bf H}_0^d = \frac{1}{2 E_\nu} {\bf diag} \left(
    0,\Delta m^2_{21}, \Delta m^2_{31} \right),
    \label{eq:evol.3}
\end{equation}
\begin{equation}
    {\bf V} =
    {\bf diag} \left( \pm \sqrt{2} G_F N_e, 0, 0 \right),
    \label{eq:evol.4}
\end{equation}
where $\vec{\nu} \equiv \left( \nu_e, \nu_\mu, \nu_\tau \right)$. We
have denoted by ${\bf H}_0^d$ the vacuum Hamiltonian, while ${\bf V}$
describes charged-current forward interactions in matter \cite{MSW}. In
Eq.~\eqref{eq:evol.4}, the sign $+$ ($-$) refers to neutrinos
(antineutrinos), $G_F$ is the Fermi coupling constant and $N_e$ is
electron number density in the Earth (note also 
that for antineutrinos, the phase $\delta$ has to be replaced with 
$-\delta$).
\begin{figure}
\centering
\includegraphics[width=0.45\textwidth]{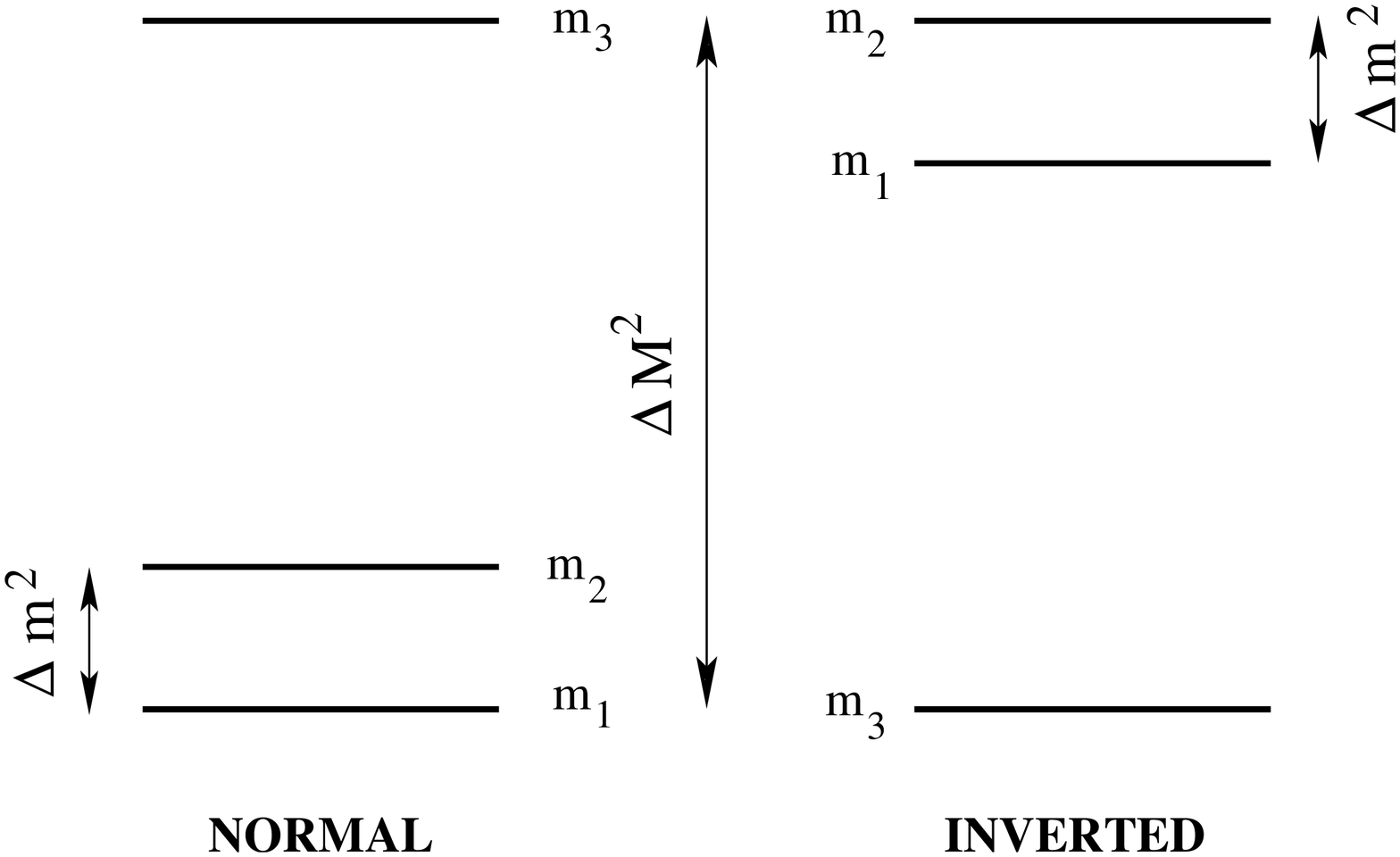}
\caption{Our convention for the mass splitting and ordering.}
\label{fig:schemes}
\end{figure}

The angles $\theta_{ij}$ can be taken without any loss of generality
to lie in the first quadrant $\theta_{ij}\in[0,\pi/2]$.  Concerning
the CP violating phase $\delta$ we chose 
the convention $0\leq \delta \leq \pi$  and  two choices of mass ordering
(See Fig.~\ref{fig:schemes}) one with $m_1\leq m_2\leq m_3$ which
we will denote as {\it Normal} and other with $m_3\leq m_1\leq
m_2$ which we will denote as {\it Inverted}
(for a recent discussion on other conventions see, for instance~\cite{zralek}).
We define as $\Delta M^2>0$ the {\it large} mass splitting 
in the problem and $\Delta
m^2>0$ the {\it small} one. In this case we can have the two mass
ordering:
\begin{eqnarray}
    \mbox{Normal:} \quad &
    \Delta M^2= \Delta m^2_{31}=m_3^2-m_1^2 & \nonumber \\
    & \Delta m^2= \Delta m^2_{21}=m_2^2-m_1^2\;\, & \label{eq:normal}
    \\
    \mbox{Inverted:} \quad &
    \Delta M^2= -\Delta m^2_{32}=m_2^2-m_3^2 &  \nonumber \\
    & \Delta m^2= \Delta m^2_{21}=m_2^2-m_1^2 \,. & \label{eq:inv1}
\end{eqnarray}
We define the {\it mass splitting hierarchy parameter}
\begin{equation}
    \alpha = \frac{\Delta m^2}{ \Delta M^2} \,,
    \label{eq:alpha}
\end{equation}
which parametrizes the departure from the one--dominant mass scale
approximation in the analysis of atmospheric and reactor neutrinos.

In this convention, for both Normal or Inverted schemes, the mixing
angles in Eq.~\eqref{eq:matrix} are such that in the one mass
dominance approximation in which $\Delta M^2$ ($\Delta m^2$)
determines the oscillation length of atmospheric (solar) neutrinos,
$\theta_{23}$ is the mixing angle relevant for atmospheric
oscillations while $\theta_{12}$ is the relevant one for solar
oscillations, and $\theta_{13}$ is mostly constrained by reactor data.
In the likely situation in which the solar solution is LMA,
$\theta_{12}$ is mainly restricted to lie in the first octant.

We will restrict ourselves to the CP conserving scenario. CP
conservation implies that the lepton phase $\delta$ is either zero or
$\pi$~\cite{Schechter:1981hw}. As we will see, for non-vanishing
$\alpha$ and $\theta_{13}$ the analysis of atmospheric neutrinos is not
exactly the same for these two possible CP conserving values of
$\delta$ and we characterize these two possibilities in terms of
$\cos\delta=\pm 1$. 

For $\alpha=\theta_{13}=0$, atmospheric neutrinos involve only
$\nu_\mu \to \nu_\tau$ conversions, and in this case there are no
matter effects, so that the solution of Eq.~\eqref{eq:evol.1} is
straightforward and the conversion probability takes the well-known
vacuum form
\begin{equation}
    P_{\mu\mu} = 1 - \sin^2 \left( 2 \theta_{23} \right)
    \sin^2 \left( \frac{\Delta M^2 L}{4 E_\nu} \right),
\end{equation}
where $L$ is the path-length traveled by neutrinos of energy $E_\nu$.

On the other hand, in the general case of three-neutrino scenario with
$\theta_{13}\neq 0$ or $\alpha\neq 0$ the presence of the matter
potentials become relevant and it requires a numerical solution of the
evolution equations in order to obtain the oscillation probabilities
for atmospheric neutrinos $P_{\alpha\beta}$, which are different for
neutrinos and anti-neutrinos because of the reversal of sign in
Eq.~\eqref{eq:evol.4}. In our calculations, we use for the matter
density profile of the Earth the approximate analytic parametrization
given in Ref.~\cite{lisi} of the PREM model of the Earth~\cite{PREM}.
\begin{figure}
\centering
    \includegraphics[width=0.45\textwidth]{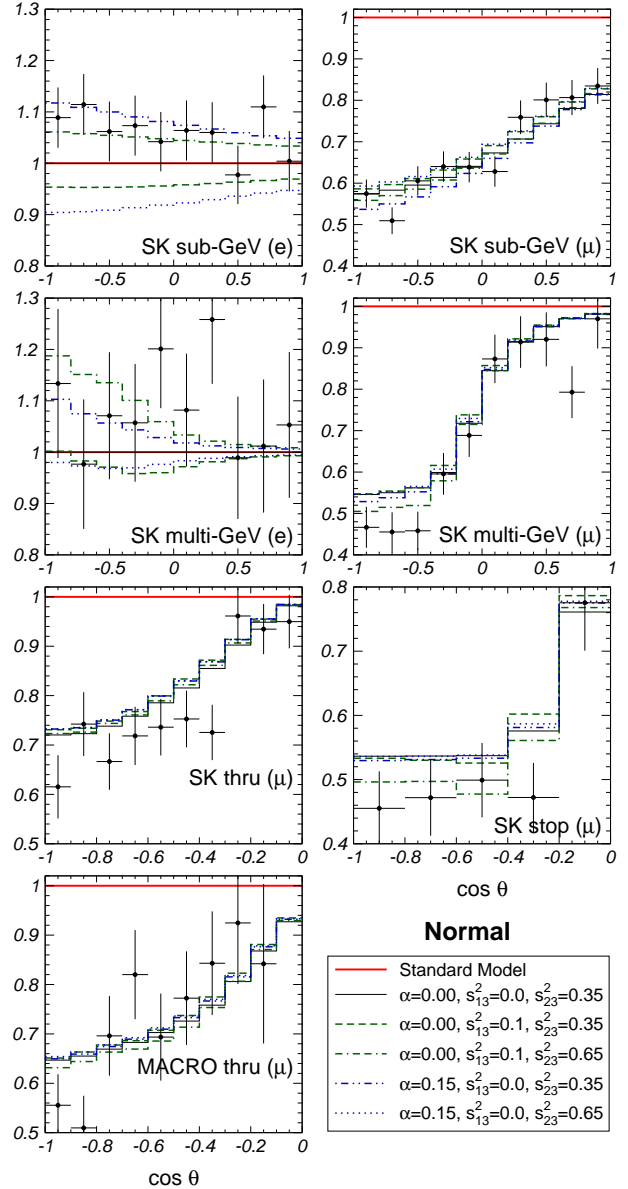}
    \caption{
      Zenith-angle distributions (normalized to the no-oscillation
      prediction) for the Super--Kamiokande $e$--like and $\mu$--like
      contained events, for the Super--Kamiokande stopping and
      through-going muon events and for Macro upgoing muons. The
      various dashed lines are the expected distributions for the
      Normal mass ordering with $\Delta M^2=3\times 10^{-3}$ eV$^2$,
      $\tan^2\theta_{12}=0.45$ and several values of
      $\sin^2\theta_{13}$ and $\sin^2\theta_{23}$ as given in the
      figure.}
\label{fig:znor}
\end{figure}

\begin{figure}
\centering
    \includegraphics[width=0.45\textwidth]{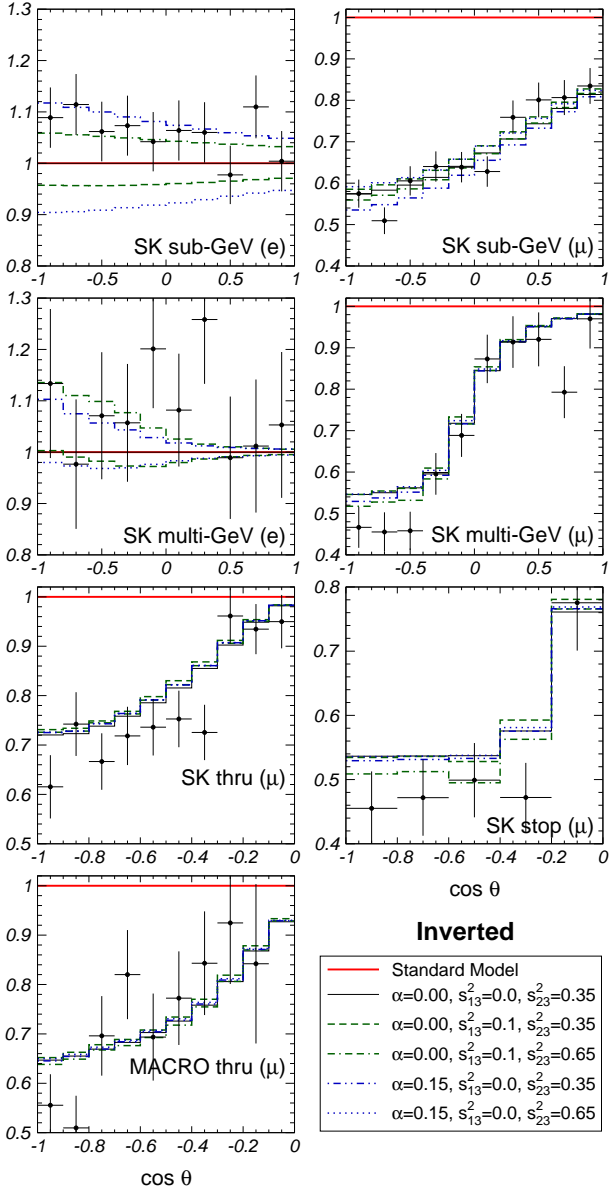}
    \caption{
      Same as Fig.~{\protect{\ref{fig:znor}}} but for Inverted mass
      ordering.}
\label{fig:zinv}
\end{figure}

In Figs.~\ref{fig:znor} and \ref{fig:zinv} we plot the angular
distribution of atmospheric $\nu_e$ and $\nu_\mu$ for non-vanishing
values of $\alpha$ or $\theta_{13}$ obtained from our numerical
calculations. As seen in these figures the main effect of a small but
non-vanishing $\alpha$ is mostly observable for sub-GeV electrons,
although some effect is also visible for multi-GeV electrons and
sub-GeV muons, and it can result either in an increase or in a
decrease of the expected number of events with respect to the
$\alpha=0$ prediction depending on whether $\theta_{23}$ is in the
first or second octant. This behaviour can be understood in terms of
the approximate analytical expressions.  
For instance for $\theta_{13}=0$ we find 
(in agreement with the results in  Ref.~\cite{smirnov1})
\begin{eqnarray}
    \frac{N_e}{N_{e0}}-1 &=&
    \overline P_{e2}
    \bar r (c_{23}^2-\frac{1}{\bar r})
    \label{eq:neal} \\
    \frac{N_\mu-N_{\mu}(\alpha=0)}{N_{\mu0}}&=& -\overline  P_{e 2}
    c_{23}^2 (c_{23}^2-\frac{1}{\bar r}) \label{eq:nmual}
\end{eqnarray}
where $N_{e0}$ and $N_{\mu0}$ are the expected number of electron and
muon-like events in the absence of oscillations in the relevant energy
and angular bin and $\bar r=N_{\mu0}/N_{e0}$.  For instance, for
sub-GeV events $\bar r\sim 2$.  Here $N_{\mu}(\alpha=0)$ is the
expected number of muon-like events for $\alpha=0$ and $\overline
P_{e2}$ is the dominant $\alpha$-dependent term in the probabilities,
averaged over energy and zenith angle. For neutrinos we have:
\begin{eqnarray}
    P_{e2} &=& \sin^22\theta_{12,m}
    \sin^2\left(\frac{\Delta m^2 \,L}{4E_\nu}
    \frac{\sin 2\theta_{12}}{\sin 2\theta_{12,m}}\right) \,, 
\label{eq:s12m}
    \\
    \sin 2\theta_{12,m} &=& \frac{\sin2\theta_{12}}
    {\sqrt{(\cos2\theta_{12}-2 E_\nu V_e/\Delta m^2)^2+
        (\sin2\theta_{12})^2} } \,,\nonumber  
\end{eqnarray}
which for $\Delta m^2\ll 2 E_\nu V_e$ reduces to:
\begin{equation}
    P_{e2}=\alpha^2\sin^22\theta_{12} \left(
    \frac{\Delta M^2}{2E V_e} \right)^2\sin^2\frac{V_e L}{2} \,.
\label{pe2alfa}
\end{equation}
According to Eqs.~\eqref{eq:neal} and \eqref{eq:nmual} the sign of the
shift in the number of predicted events with respect to the results in
the one mass scale dominance approximation is opposite for electron
and muon-like events and it depends on the factor
$c_{23}^2-\frac{1}{\bar r}\sim c_{23}^2-0.5$. So for $\theta_{23}$ in
the first octant, $c_{23}^2>0.5$, there is an increase (decrease) in
the number of electron (muon) events as compared to the $\alpha=0$
case. For $\theta_{23}$ in the second octant the opposite holds. We
also see that the net shift is larger for electron events than for
muon events by a factor $c_{23}^2/\bar r$. Notice that, despite 
Eq.~(\ref{pe2alfa}) looks order $\alpha^2$, its numerical value 
for sub-GeV electrons is large due to the factor $\Delta M^2/(2 E V_e)$ 
as can be seen from the figures. 
At higher energies, for up-going muons the effect is negligible.

For the sake of comparison we also show in the figures the behaviour
with non-vanishing value of $\theta_{13}$ in the one mass scale
dominance approximation. As seen in the figure the effect is most
important for the electron events and can be understood as follows.
For the case of constant matter density the expected flux of $\nu_e$
events in the one mass scale dominance approximation we find 
\begin{equation}
    \frac{N_e}{N_{e0}}-1 = 
\overline P_{e\mu} \bar r (s_{23}^2-\frac{1}{\bar r})
\end{equation}
where
\begin{eqnarray}
    P_{e\mu}&=& 4 s^2_{13,m} c^2_{13,m} \sin^2\left(
    \frac{\Delta M^2 \,L}{4E_\nu}
    \frac{\sin 2\theta_{13}}{\sin 2\theta_{13,m}} \right) \,,
    \label{eq:s13m} \\
    \sin2\theta_{13,m}&=&
    \frac{\sin2\theta_{13}}
    {\sqrt{(\cos2\theta_{13}\mp
        2 E_\nu V_e/\Delta M^2)^2+(\sin2\theta_{13})^2} } 
\nonumber
\end{eqnarray}
and the $-$ (+) sign applies for the Normal (Inverted) case    
(similar expression is presented, for instance, in the last article 
in Ref.~\cite{2atm} and in Ref.~\cite{atmpetcov}). 
So for $\theta_{23}$ in the first octant ($s_{23}^2<0.5$) there is a decrease
in the number of electron events as compared to the $\theta_{13}$
case.  For sub-GeV events, the matter term in Eq.~\eqref{eq:s13m} can be
neglected and the effect of a non-vanishing $\theta_{13}$ is the same
for Normal and Inverted ordering.  For multi-GeV and upgoing muon
events matter
effects start playing a role and the effect becomes slightly larger
for the Normal case where the matter enhancement is in the neutrino
channel.

The situation becomes more involved when both $\alpha$ and
$\theta_{13}$ are different from zero. For instance, in lowest order
in $\alpha$ $s_{13}$ the expected number of sub-GeV $\nu_e$ events is
(after averaging the $\Delta M^2 L/E$ oscillations)
\begin{eqnarray}
    \frac{N_e}{N_{e0}}-1 
    &=& \overline{P}_{e2} \, \bar{r} \,(c_{23}^2-\frac{1}{\bar r}) 
    + \overline{P}_{e\mu} \, \bar r\, (s_{23}^2-\frac{1}{\bar r})
    \label{eq:nefull}    
    \\
    &+& \frac{\bar r}{2}\cos\delta
    \sin 2\theta_{13}\sin 2\theta_{23}\sin 2\overline \theta_{12,m}
    \cos 2\overline \theta_{12,m} \nonumber \\
&& \overline{\sin^2\left(\frac{\Delta m^2 \,L}{4E_\nu}
    \frac{\sin 2\theta_{12}}{\sin 2\theta_{12,m}} \right)}
\nonumber 
\end{eqnarray}
(this expression is in agreement with the results in Ref.~\cite{smirnov2}). 
From this equation we see that the {\it interference} term (the 
third term in the right hand side) can have either sign depending on
$\cos\delta$.  
It also changes sign depending on whether the $\Delta
m^2$ oscillations are above ($\Delta m^2 \cos 2\theta_{12}
> 2 E_\nu V_e$) of below
($\Delta m^2\cos 2\theta_{12} < 2 E_\nu V_e$) 
the resonance.  For very small $\alpha$ 
($\Delta m^2 \ll 2 E_\nu V_e)$ the interference term is proportional 
to $\alpha$ and it
also changes sign for neutrinos and antineutrinos.  In summary the
effect of non-vanishing $\theta_{13}$ and $\alpha$ in the expected
number and distribution of atmospheric neutrino events can have
opposite signs and this can lead to a partial cancellation between
both contributions. This results into a loss of sensitivity of the
analysis to both parameters.

To analyze the CHOOZ constraints we need to evaluate the survival
probability for ${\bar \nu}_e$ of average energy $E\sim$ few MeV at a
distance of $L\sim 1$~Km. For these values of energy and distance, one
can safely neglect Earth matter effects. The survival probability
takes the analytical form:
\begin{eqnarray}
    P_{ee}^{\rm CHOOZ}
    &=& 1-\cos^4\theta_{13}\sin^22\theta_{12}
    \sin^2\left(\frac{\Delta m^2_{21} L}{4 E_\nu} \right) \nonumber
    \\
    & & -\sin^22\theta_{13}\left[\cos^2\theta_{12}
    \sin^2\left(\frac
    {\Delta m^2_{31} L}{4 E_\nu} \right) \right. \nonumber \\
    && \left.+ \sin^2\theta_{12} \sin^2 \left( 
    \frac{\Delta m^2_{32} L}{4 E_\nu} \right) \right] \label{eq:pchooz}
    \\ 
    &\simeq& 1-\sin^22\theta_{13}\sin^2\left(
    \frac{\Delta M^2 L}{4E_\nu}\right), \nonumber
\end{eqnarray}
where the second equality holds under the approximation $\Delta m^2\ll
E_\nu/L$ which can only be safely made for $\Delta m^2\leq 3\times
10^{-4}~$eV$^2$. Eq.~\eqref{eq:pchooz} is valid for both Normal 
and Inverted ordering with the identifications in 
Eq.\eqref{eq:normal}  and Eq.\eqref{eq:inv1} respectively. 
It results that the probability for Normal and Inverted schemes  is
the same with the exchange 
$\sin^2\theta_{12}\leftrightarrow\cos^2\theta_{12}$. 
Thus in general the analysis of the CHOOZ reactor data
involves four oscillation parameters: $\Delta M^2$, $\theta_{13}$,
$\Delta m^2$, and $\theta_{12}$. From Eq.~\eqref{eq:pchooz} we see
that for a given value of $\theta_{12}$ and $\Delta M^2$ the effect of
a non-vanishing value of either $\theta_{13}$ or $\Delta m^2$ is the
decrease of the survival probability.

\section{Atmospheric Neutrino Analysis}
\label{sec:analatmos}

In our statistical analysis of the atmospheric neutrino events we use
all the samples of SK data: $e$-like and $\mu$-like samples of sub-
and multi-GeV~\cite{skatmlast} data, each given as a 10-bin
zenith-angle distribution
, and upgoing muon data including
the stopping (5 bins in zenith angle) and through-going (10 angular
bins) muon fluxes. We have also included the latest MACRO~\cite{MACRO}
upgoing muon samples, with 10 angular bins. So we have a total of 65
independent inputs.

For details on the statistical analysis applied to the different
observables, we refer to the first reference in Refs.~\cite{2atm} and
\cite{3ours}. As discussed in the previous section, the analysis of
the atmospheric neutrino data for three neutrino oscillations with two
mass scales involves six parameters: two mass differences, three
mixing angles and one CP phase. 
Our aim is to study the modification on the resulting allowed ranges of the
parameters $\Delta M^2$, $\sin^2\theta_{23}$ and $\sin^2\theta_{13}$
due to the deviations from the one--dominant mass scale approximation,
{\it i.e.}\ for $\Delta m^2\neq 0$ (or equivalently for non-vanishing
values of mass splitting hierarchy parameter $\alpha$). 
In what follows, for the sake of
simplicity, we will restrict ourselves to the CP conserving scenario
but we will distinguish the two possible CP conserving values of
$\delta$ and we characterize these two possibilities in terms of
$\cos\delta=\pm 1$. We will show the results for Normal and Inverted
schemes. Furthermore in most of our study we will keep the mixing
angle $\theta_{12}$ to be within the LMA range favoured in the 
global analysis of solar neutrino data by choosing a characteristic 
value $\tan^2\theta_{12}=0.45$~\cite{2solour,2solother}.
We have repeated our analysis for different values of
$\theta_{12}$ and we have found that the 
maximum effect due to the variation of $\theta_{12}$ is a shift 
on $\Delta\chi^2\sim 1$ and it is therefore
unobservable. Furthermore we have verified that the atmospheric data 
analysis does not provide enough precision to test the possibility 
of non-vanishing CP violation. 

\begin{figure}
\centering
\includegraphics[width=0.45\textwidth]{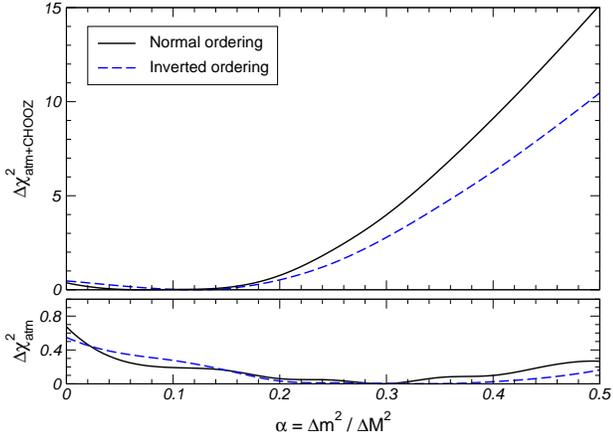}
    \caption{\label{fig:chialpha}%
      Dependence of the $\Delta\chi^2_{\rm atm}$ and
      $\Delta\chi^2_{\rm atm+CHOOZ}$ functions on the mass splitting
      parameter $\alpha$ for $\tan^2\theta_{12}=0.45$, for the analysis
      of atmospheric neutrinos alone (lower panel) and also in
      combination with the CHOOZ reactor data (upper panel).}
\end{figure}

\begin{figure*}
\centering
    \includegraphics[height=0.45\textheight]{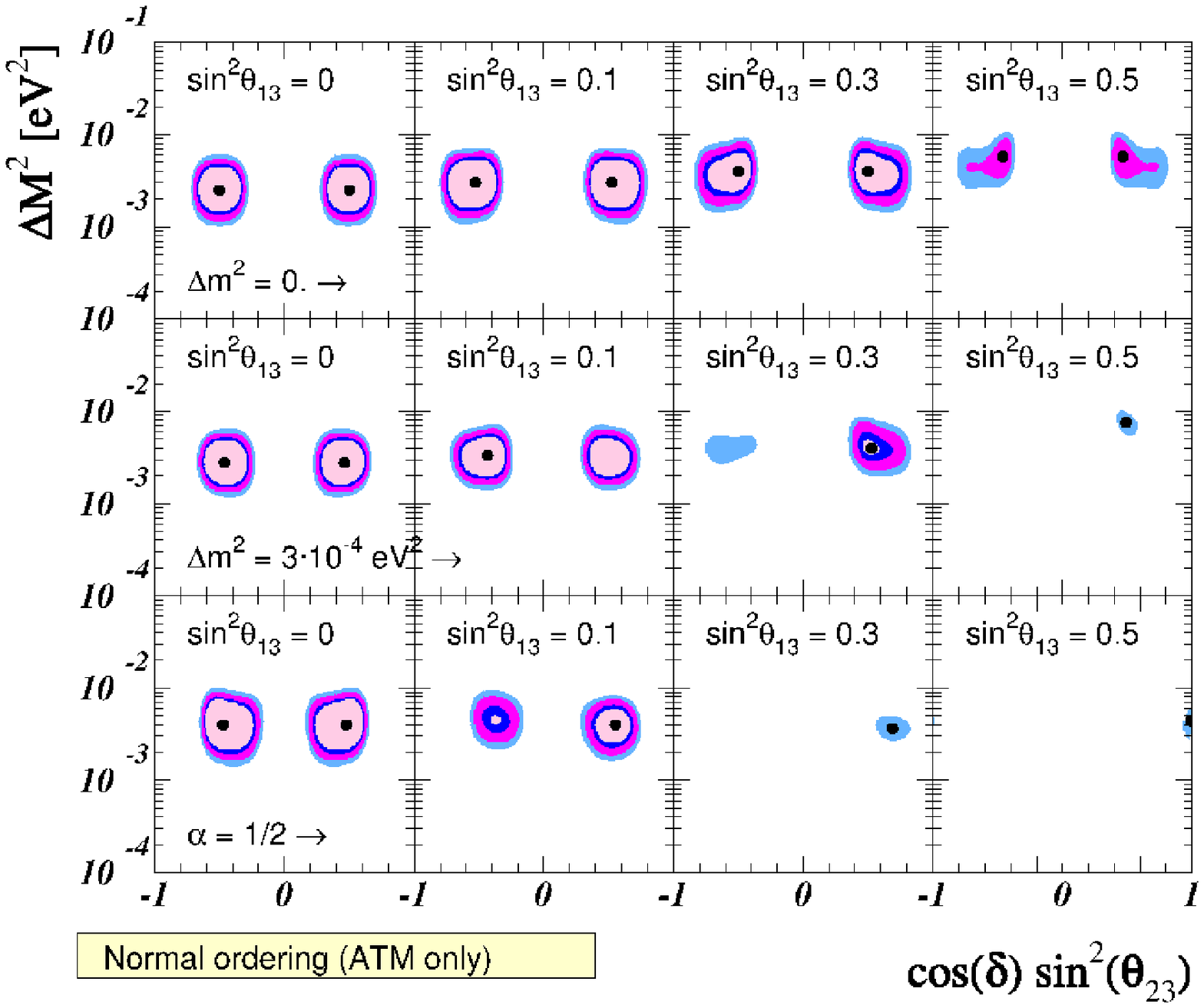} \\[5mm]
    \includegraphics[height=0.45\textheight]{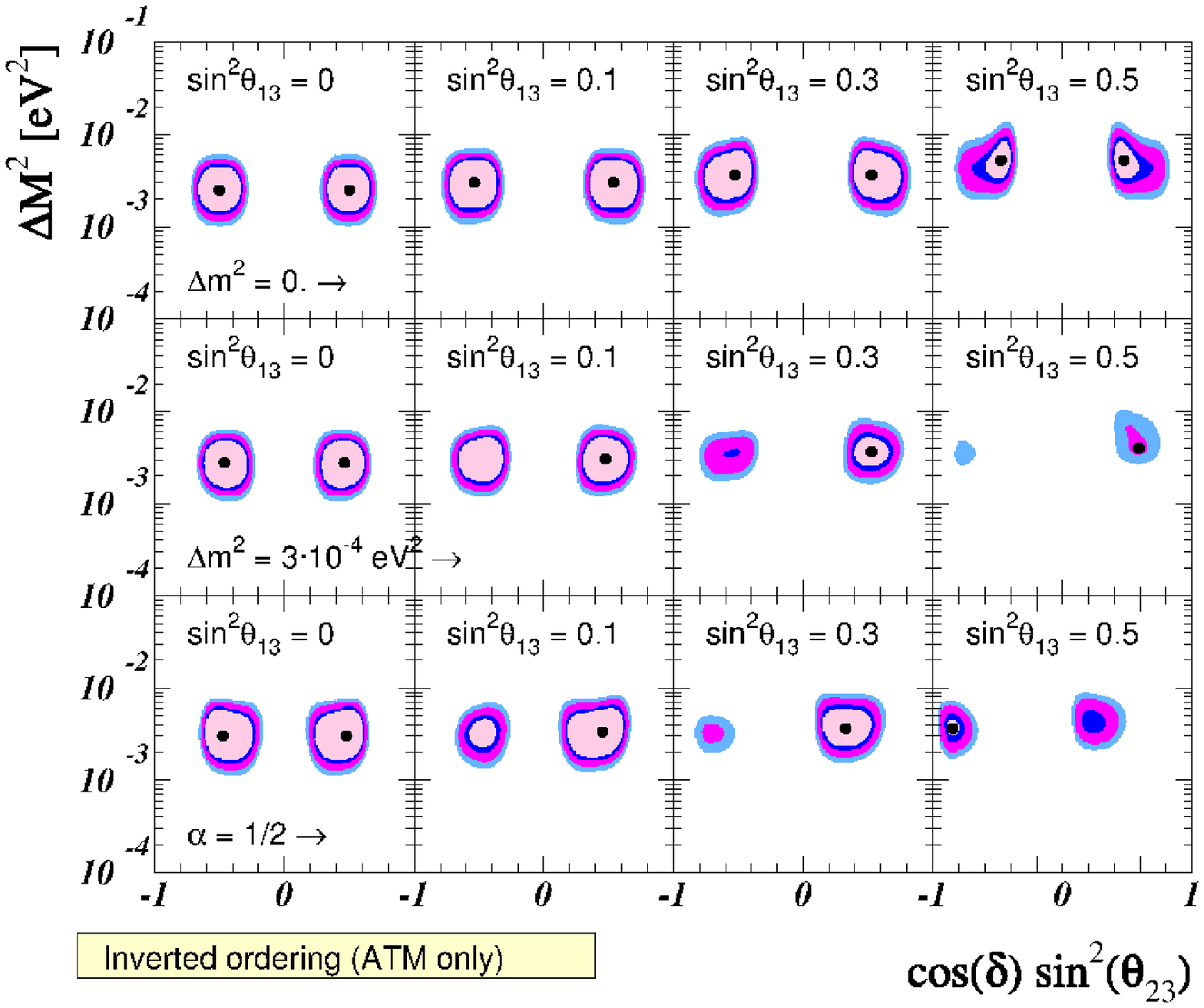}
    \caption{
      90\%, 95\%, 99\% and $3\sigma$ (4 d.o.f.) allowed regions in the
      $(\sin^2\theta_{23}, \Delta M^2)$ plane, for different values of
      $\sin^2\theta_{13}$ and $\Delta m^2$, from the analysis of the
      atmospheric neutrino data. The global minimum used to define the
      allowed regions is given in Eq.~\eqref{eq:minatm}; the local
      minima are marked with a dot.}
\label{fig:alpha}%
\end{figure*}

\begin{figure*}
\centering
    \includegraphics[width=0.9\textwidth]{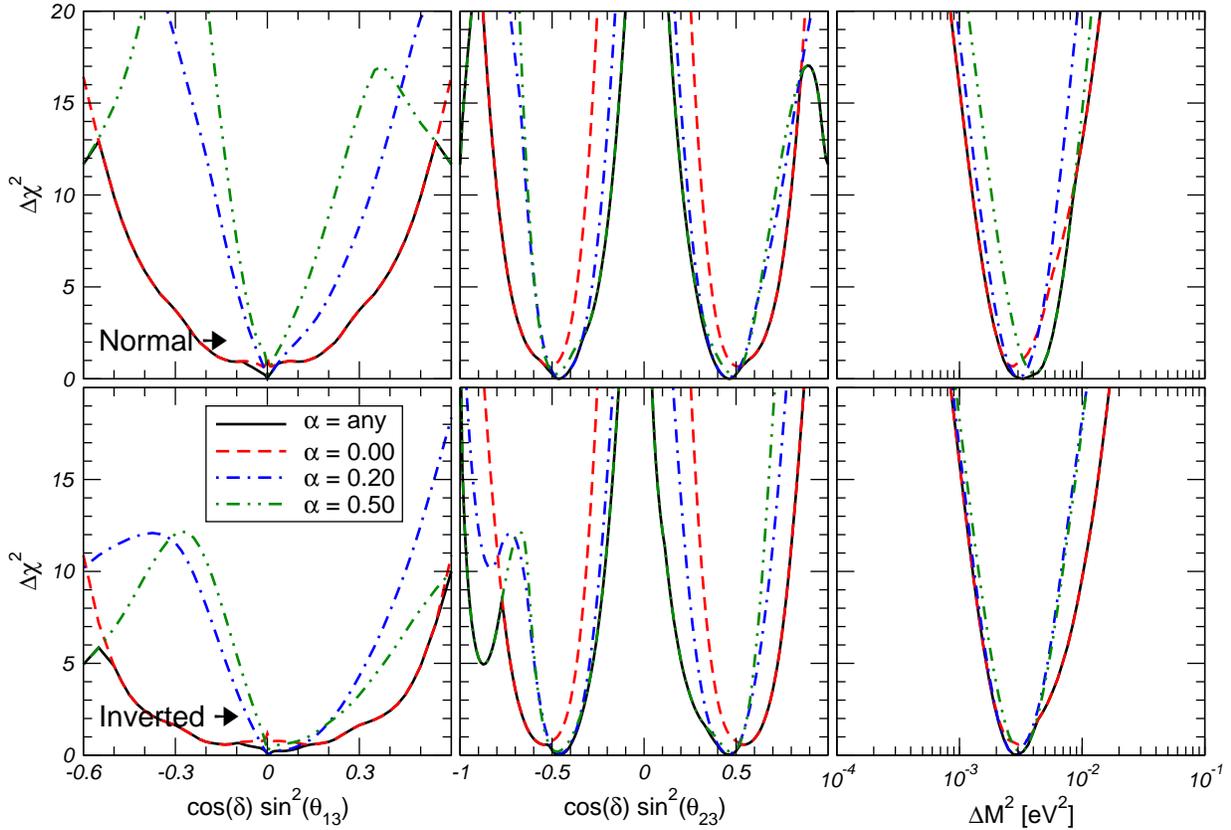}
    \caption{
      Dependence of the $\Delta \chi^2_{\rm atm}$ function on the
      mixing angles $\cos\delta\sin^2\theta_{23}$ and
      $\cos\delta\sin^2\theta_{13}$ and on the large mass scale
      $\Delta M^2$, for different values of $\alpha$ and for the
      Normal (upper panels) and Inverted (lower panels) cases. See
      text for details.}
\label{fig:chisnc}%
\end{figure*}

\begin{figure}
\centering
    \includegraphics[width=0.45\textwidth]{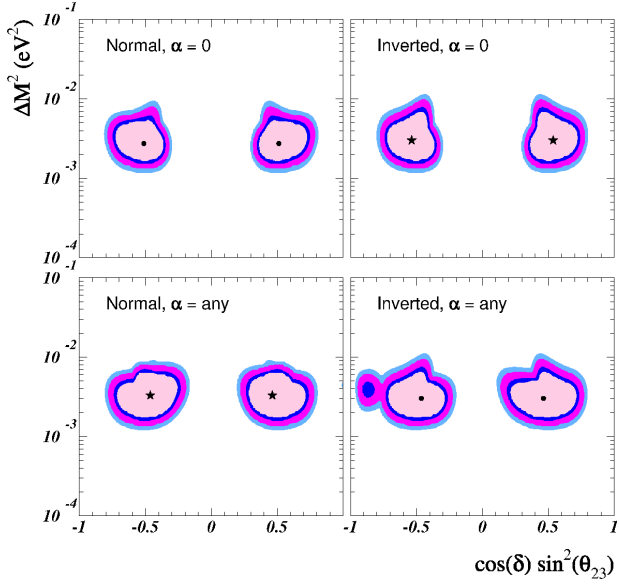}
    \caption{
      90\%, 95\%, 99\% and $3\sigma$ (2 d.o.f.) allowed regions in the
      $(\cos\delta\sin^2\theta_{23}, \Delta M^2)$ plane from the
      analysis of the atmospheric neutrino data, for the Normal (left
      panels) and Inverted (right panels) cases, for
      $\tan^2\theta_{12}=0.45$ and for arbitrary values of
      $\theta_{13}$ and $\alpha$ (lower panels). See text for details.
      The upper panels correspond to the case $\alpha=0$. The best fit
      point in each case is marked with a star. The local minima are
      marked with a dot.}
\label{fig:globalnc}
\end{figure}

We first present the results of the allowed parameters for the global
combination of atmospheric observables. Notice that since the
parameter space we study is four-dimensional the allowed regions for a
given CL are defined as the set of points satisfying the condition for
four degrees of freedom (d.o.f.)
\begin{equation}
    \chi^2_{\rm atm}(\Delta M^2, \theta_{23},\theta_{13}, \Delta m^2)
    -\chi^2_{\rm atm,min}\leq \Delta\chi^2(\mbox{CL, 4~d.o.f.})
\end{equation}
where $\Delta\chi^2(\mbox{CL, 4~d.o.f.}) = 7.78$, $9.49$, $13.3$ and
$16.25$ for CL~= $90\%$, $95\%$, $99\%$ and $99.73\% \equiv 3\sigma$
respectively, and $\chi^2_{\rm atm, min}$ is the global minimum in the
four-dimensional space. The best fit point used to define the allowed
parameter space is found to be:
\begin{eqnarray}
    \Delta M^2 &=& 3.3\times 10^{-3}~{\rm eV}^2 \nonumber
    \\
    \sin^2\theta_{23} &=& 0.46\nonumber 
    \\
    \sin^2\theta_{13} &=& 0. \label{eq:minatm} 
    \\
    \Delta m^2 &=& 1.0\times 10^{-3}~{\rm eV}^2 \quad
    (\alpha=0.30) \nonumber
    \\
    \chi^2_{\rm atm, min} &=& 39.0 \nonumber
\end{eqnarray}
(for 61 d.o.f.) and it corresponds to Normal ordering 
although the difference with the Inverted ordering
($\Delta\chi^2=0.1$) is not statistically significant \footnote{The
careful reader may notice that the $\chi^2$ per d.o.f. seems {\sl too good}. 
This was already the case for the previous SuperKamiokande data sample
and it is partly due to the very good agreement of the multi-GeV 
electron distributions with their no-oscillation expectations}.
The point given in Eq.~\eqref{eq:minatm} is the global minimum used in
the construction of the $\Delta\chi^2_{\rm atm}$ function shown in
Figs.~\ref{fig:chialpha} and \ref{fig:chisnc}, of the allowed
parameter space shown in Fig.~\ref{fig:alpha} and in the lower panels
of Fig.~\ref{fig:globalnc}, and of the ranges in
Eq.~\eqref{eq:atmparam}.

This result can be compared with the best fit point obtained in the
one--dominant mass scale approximation $\alpha=0$
\begin{eqnarray}
    \Delta M^2 &=& 3.0\times 10^{-3}~{\rm eV}^2 \nonumber
    \\
    \sin^2\theta_{23} &=& 0.54 \nonumber
    \\
    \sin^2\theta_{13} &=& 0.14 \label{eq:minatm0} 
    \\
    \chi^2_{\rm atm, min} &=& 39.6 \nonumber
\end{eqnarray}
(for 62 d.o.f., one more than in the $\alpha$-unconstrained case) 
which is independent of the choice $\cos\delta=\pm 1$ and corresponds
to Inverted schemes. Notice that this is the minimum used to obtain
the allowed parameter space in the one-dominant mass scale
approximation [the upper panels in Fig.~\ref{fig:globalnc} and the
ranges in Eq.~\eqref{eq:atmparam0}], since in this case we are fixing
{\sl a priori} $\alpha=0$.

In summary, we find that allowing for a non-zero value of $\alpha$
very mildly improves the quality of the global fit. This result is
driven by the better description of the sub-GeV data which is
attainable for a non-zero $\alpha$ value, and drives the best fit
point to the first octant of the mixing angle $\theta_{23}$ for which
the expected number of sub-GeV electrons is larger as compared to the
pure $\nu_\mu\rightarrow\nu_\tau$ scenario, as illustrated in
Figs.~\ref{fig:znor} and~\ref{fig:zinv}. 
We find, however, that the
analysis of the atmospheric neutrino data does not show a strong
dependence on large $\alpha$ values. In the lower panel of
Fig.~\ref{fig:chialpha} we show the dependence of $\Delta\chi^2_{\rm
atm}$ on $\alpha$.  In this plot all the neutrino oscillation
parameters which are not displayed have been ``integrated out'', {\it
i.e.}\ the $\Delta\chi^2_{\rm atm}$ function is minimized with respect
to all the non-displayed variables. From this figure we see that the
fit to atmospheric neutrinos is only weakly  sensitive to the value of
$\alpha$.

In Fig.~\ref{fig:alpha} we present sections of
the allowed volume in the plane $(\cos\delta\sin^2\theta_{23},~\Delta
M^2)$ for different values of $\sin^2\theta_{13}$ and 
for values of $\Delta m^2=0$ (first row) and 
$\Delta m^2=3\times 10^{-4}~$eV$^2$ (second row) which is the (maximum
allowed value by the present analysis of solar neutrino
data including the latest 1496 days of SK and and day-night
spectrum of SNO data~\cite{2solour}. 
For illustration we also show 
the corresponding regions for the ``democratic'' scenario 
$\alpha=0.5$ ($\Delta m^2=\Delta M^2/2$). 
We display the corresponding sections for the Normal and
Inverted schemes. 

Comparing the sections in Fig.~\ref{fig:alpha} for $\alpha=0$
with the corresponding sections for non vanishing $\alpha$ values we
find that substantial differences appear although mainly for 
large values of $\theta_{13}$. However
from these figures one also realizes that even for large values of
$\alpha$ the allowed region does not extend to a very different range
of $\Delta M^2$. Conversely, the mixing angles $\theta_{23}$ and
$\theta_{13}$ can become less constrained when the case $\alpha\neq 0$
is considered.

To further quantify these effects we plot in Fig.~\ref{fig:chisnc} the
dependence of $\Delta\chi^2_{\rm atm}$ on $\Delta M^2$, $\theta_{23}$
and $\theta_{13}$, respectively, for different values of $\alpha$,
after minimizing with respect to all the non-displayed variables. From
these figures we can read the $3\sigma$ allowed ranges for the
different parameters (1 d.o.f.): 
\begin{itemize}
\item[$\bullet$] For arbitrary $\alpha$
\begin{equation}
    \begin{tabular}{c@{\hspace{10mm}}c}
        Normal & Inverted \\
        $1.3\leq \frac{\Delta M^2}{10^{-3}~{\rm eV^2}}\leq 8.1$ &
        $1.2\leq \frac{\Delta M^2}{10^{-3}~{\rm eV^2}}\leq 9.6$ 
        \\
      \multicolumn{2}{c}{$\cos\delta=1$}    \\
        $0.22\leq \sin^2\theta_{23}\leq 0.79$ &
        $0.14\leq \sin^2\theta_{23}\leq 0.78$  \\
        $\sin^2\theta_{13}\leq 0.48$ &
        $\sin^2\theta_{13}\leq 0.58$     \\
 \multicolumn{2}{c}{$\cos\delta=-1$}  \\
        $0.19\leq \sin^2\theta_{23}\leq 0.79$ &
        $0.22\leq \sin^2\theta_{23}\leq 0.95$  \\
        $\sin^2\theta_{13}\leq 0.48$ &
        $\sin^2\theta_{13}\leq 1$
    \end{tabular}
    \label{eq:atmparam}
\end{equation}
\item[$\bullet$] For  $\alpha=0$ (no dependence on $\cos\delta$)
\begin{equation}
    \begin{tabular}{c@{\hspace{10mm}}c}
        Normal & Inverted \\
        $1.3\leq \frac{\Delta M^2}{10^{-3}~{\rm eV^2}}\leq 8.1$ &
        $1.3\leq \frac{\Delta M^2}{10^{-3}~{\rm eV^2}}\leq 10.0$
        \\
        $0.32\leq \sin^2\theta_{23}\leq 0.79$ &
        $0.32\leq \sin^2\theta_{23}\leq 0.78$ \\
        $\sin^2\theta_{13}\leq 0.49$ &
        $\sin^2\theta_{13}\leq 0.58$ \\
    \end{tabular}
    \label{eq:atmparam0}
\end{equation}
\end{itemize}

Comparing the ranges in Eqs.~\eqref{eq:atmparam} and
\eqref{eq:atmparam0} we see that the parameter which is less sensitive
to the departure from the one mass scale dominance approximation is
$\Delta M^2$, while $\sin^2\theta_{13}$ is the mostly affected, in
particular for the Inverted scheme for which no upper bound on
$\sin^2\theta_{13}$ is derived from the analysis. The careful reader
may notice that for the Normal ordering, the bound on $\theta_{13}$
for arbitrary $\alpha$ can be stronger than for $\alpha=0$. This is
due to the fact that the ranges in Eqs.~\eqref{eq:atmparam} and
\eqref{eq:atmparam0} are defined in terms of $3\sigma$ shifts in the
$\chi^2$ function with respect to the minima in Eqs.~\eqref{eq:minatm}
and \eqref{eq:minatm0} respectively [see explanation below 
Eq.~\eqref{eq:minatm0}].

Finally we show in Fig.~\ref{fig:globalnc} the 2-dimensional allowed 
regions in $(\cos\delta\sin^2\theta_{23}, \Delta M^2)$ from the
analysis of the atmospheric neutrino data independently of the values
of $\alpha$ and $\theta_{13}$.  In constructing these regions for each
value of $\Delta M^2$ and $\cos\delta\sin^2\theta_{23}$ we have
minimized on the oscillation parameters $\Delta m^2$ and $\theta_{13}$
so the they are defined in terms of $\Delta\chi^2$ for 2 d.o.f.\
($\Delta\chi^2=4.61$, $5.99$, $9.21$, $11.8$ for 90\%, 95\%, 99\% CL
and $3\sigma$ respectively). For the sake of comparison we show in the
figure the corresponding regions for $\alpha=0$. From the figure we
see that the differences are larger for the Inverted scheme.

\section{Analysis of CHOOZ Data}
\label{sec:analchooz}

The CHOOZ experiment~\cite{CHOOZ} searched for disappearance of
$\bar{\nu}_e$ produced in a power station with two pressurized-water
nuclear reactors with a total thermal power of $8.5$~GW (thermal).  At
the detector, located at $L\simeq 1$~Km from the reactors, the
$\bar{\nu}_e$ reaction signature is the delayed coincidence between
the prompt ${\rm e^+}$ signal and the signal due to the neutron
capture in the Gd-loaded scintillator.  Their measured vs.\ expected
ratio, averaged over the neutrino energy spectrum is
\begin{equation}
    R = 1.01 \pm 2.8 \,\% ({\rm stat}) \pm 2.7 \,\% ({\rm syst}) \,.
    \label{eq:rchooz}
\end{equation}
Thus no evidence was found for a deficit of measured vs.\ expected
neutrino interactions, and they derive from the data exclusion plots
in the plane of the oscillation parameters $(\Delta m^2,\, \sin^2 2
\theta)$ in the simple two-neutrino oscillation scheme. At 90\% CL
they exclude the region given approximately by $\Delta m^2 > 7 \cdot
10^{-4}~$eV$^2$ for maximum mixing, and by $\sin^2(2\theta) > 0.10$
for large $\Delta m^2$. Similar searches have been performed at the Palo
Verde Reactor Experiment \cite{paloverde} leading to slightly weaker
bounds.

\begin{figure}
\centering
\includegraphics[width=0.45\textwidth]{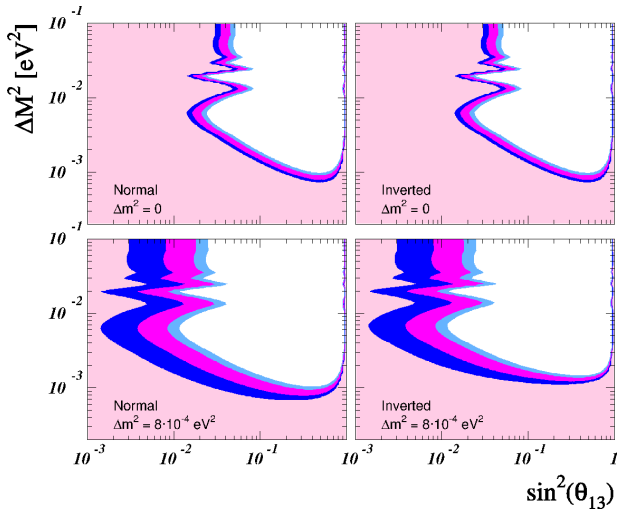}
    \caption{
      90\%, 95\%, 99\% and $3\sigma$ (2 d.o.f.) allowed regions from
      the analysis of the CHOOZ reactor data in the
      $(\sin^2\theta_{13},\, \Delta M^2)$ plane for different values of
      $\Delta m^2$ ($\tan^2\theta_{12}=0.45$), for the Normal (left
      panels) and Inverted (right panels) cases.}
\label{fig:chooz}
\end{figure}

In order to combine the CHOOZ bound with the results from our analysis
of atmospheric neutrino data in the framework of three-neutrino mixing
we have first performed our own analysis of the CHOOZ data. Using as
experimental input their measured ratio \eqref{eq:rchooz}~\cite{CHOOZ}
and comparing it with the theoretical expectations we define the
$\chi^2_{\rm CHOOZ}$ function. 
We verified that with our $\chi^2_{\rm CHOOZ}$ function
and using the statistical criteria for two degrees of freedom we
reproduce the excluded regions given in Ref.~\cite{CHOOZ} 
as can be seen in the upper row of Fig.~\ref{fig:chooz} \footnote{   
For the sake of simplicity we chose not to include the energy dependence 
of the CHOOZ data  which adds very little to the knowledge of the parameter 
space as can be seen by comparing our results in Fig.~\ref{fig:chooz} 
with those of the CHOOZ collaboration~\cite{CHOOZ} or the corresponding
ones in the analysis of Ref.~\cite{choozpetcov}.}
As discussed in
Sec.~\ref{sec:parameters} for the analysis of the reactor data the
relevant oscillation probability depends in general on the four
parameters $\theta_{12}$, $\Delta m^2$, $\theta_{13}$, and $\Delta
M^2$. In Fig.~\ref{fig:chooz} we show the excluded regions at 90, 95
and 99\% CL and $3\sigma$ in the $(\sin^2\theta_{13},\Delta M^2)$
plane from our analysis of the CHOOZ data for several values of
$\Delta m^2$ and $\tan^2\theta_{12}=0.45$; for the sake of comparison
with the 2-family analysis we have defined the allowed regions for 2
d.o.f.\ ($\Delta\chi^2_{\rm CHOOZ}=4.61$, $5.99$, $9.21$, $11.83$
respectively). In the left (right) panel we show the results for the
Normal (Inverted) scheme. We see that the presence of a non-vanishing
value of $\Delta m^2$ results into a slightly smaller allowed range of
$(\Delta M^2,\sin^2\theta_{13})$. For the chosen value of
$\tan^2\theta_{12}$ the reduction for smaller values of $\Delta M^2$
is slightly more significant for the Normal than for the Inverted
scheme as also shown in Ref.~\cite{choozpetcov}. This can be easily
understood from the expression of the survival probability: from 
Eq.~\eqref{eq:pchooz}, we get
\begin{eqnarray}
    P_{ee,\rm NOR}^{\rm CHOOZ}-P_{ee,\rm INV}^{\rm CHOOZ}&=&
    -\sin^22\theta_{13}(\cos^2\theta_{12}-\sin^2\theta_{12}) \nonumber
    \\
    \left[\sin^2\left(
    \frac{\Delta M^2 L}{4 E_\nu}\right)\right. &-& \left. \sin^2\left(
    \frac{(\Delta M^2 - \Delta m^2) L}{4 E_\nu}\right)\right] 
    \label{eq:choozdif}
\end{eqnarray}
Thus for $\theta_{12}\leq \frac{\pi}{4}$ 
the survival probability is 
smaller for the Normal ordering
than for the Inverted one, which leads to the stronger constraint. For
$\Delta M^2\gg \Delta m^2$, Eq.~\eqref{eq:choozdif} vanishes and the
excluded regions in the two schemes become indistinguishable.

\section{Combined Analysis}
\label{sec:combined}

\begin{figure*}
\centering
    \includegraphics[width=0.85\textwidth]{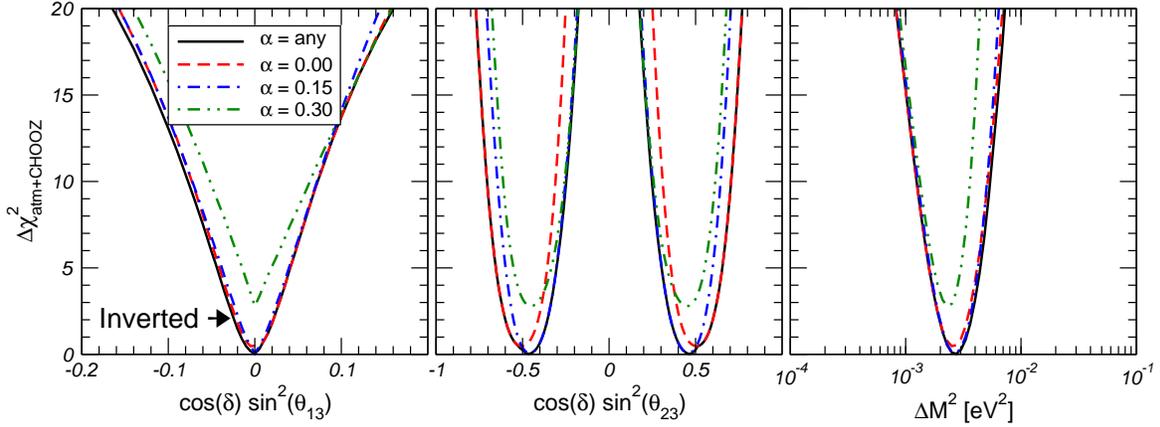}
    \caption{
      Dependence of the $\Delta \chi^2_{\rm atm+CHOOZ}$ function on
      the mixing angles $\cos\delta\sin^2\theta_{23}$ and
      $\cos\delta\sin^2\theta_{13}$ and on the large mass scale
      $\Delta M^2$, for different values of $\alpha$ and for the
      Inverted case. See text for details.}
\label{fig:chiswc}
\end{figure*}

\begin{figure} 
\centering
    \includegraphics[width=0.45\textwidth]{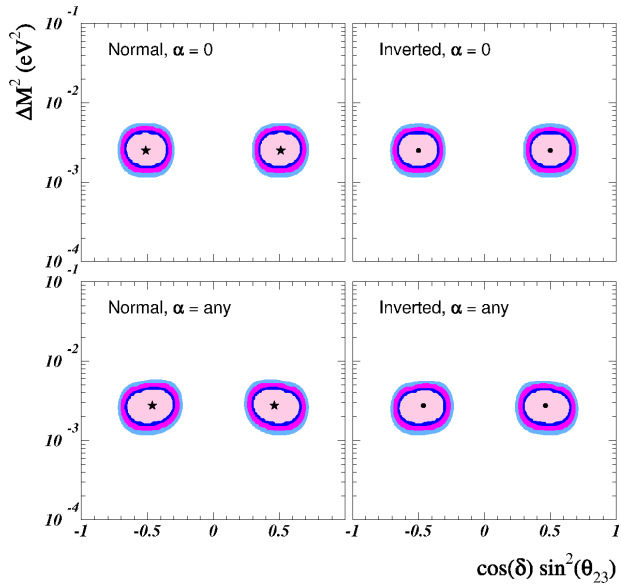}
    \caption{
      90\%, 95\%, 99\% and $3\sigma$ (2 d.o.f.) allowed regions in the
      $(\cos\delta\sin^2\theta_{23}, \Delta M^2)$ plane, from the
      analysis of the atmospheric and CHOOZ neutrino data with
      $\tan^2\theta_{12}=0.45$, for the Normal (left panels) and
      Inverted (right panels) cases and for arbitrary values of
      $\theta_{13}$ and $\alpha$ (lower panels). See text for details.
      The upper panels correspond to the case $\alpha=0$. The best fit
      point in each case is marked with a star. The local minima are
      marked with a dot.}
\label{fig:globalwc}%
\end{figure}

We now describe the effect of including the CHOOZ reactor data
together with the atmospheric data samples in a combined 3-neutrino
$\chi^2$ analysis. The results of this analysis are summarized in the
upper panel of Fig.~\ref{fig:chialpha} and in
Figs.~\ref{fig:chiswc}--\ref{fig:globalwc}. As in
Sec.~\ref{sec:analatmos}, in most of results shown here we fix the
mixing angle $\tan^2\theta_{12}=0.45$ and study how the allowed ranges of
the parameters $\Delta M^2$ , $\sin^2\theta_{23}$ and
$\sin^2\theta_{13}$ depend on $\alpha$.

We find that, in this case, the best fit point for the combined
analysis of atmospheric and CHOOZ data is practically insensitive to
the choice of Normal or Inverted schemes and:
\begin{eqnarray}
    \Delta M^2 &=& 2.8\times 10^{-3}~{\rm eV^2} \nonumber
    \\
    \sin^2\theta_{23} &=& 0.46 \nonumber 
    \\
    \sin^2\theta_{13} &=& 0.   \label{eq:miniatmcho}
    \\
    \Delta m^2 &=& 2.8\times 10^{-4}~{\rm eV^2} \quad
    (\alpha=0.1) \nonumber
    \\
    \chi^2_{\rm atm+CHOOZ, min}&=& 39.8 \nonumber
\end{eqnarray}
for 62 d.o.f.\ and $\cos\delta=\pm 1$. Notice than in our analysis the CHOOZ
data adds only one data point.
Note that the point given in Eq.~\eqref{eq:miniatmcho} is the global
minimum used in the construction of the $\Delta\chi^2_{\rm atm+CHOOZ}$
function shown in Figs.~\ref{fig:chialpha} and \ref{fig:chiswc}, of
the allowed regions in the lower panels of Fig.~\ref{fig:globalwc},
and in the ranges in Eq.~\eqref{eq:atmchparam}.

For $\alpha=0$ the best fit point is at:
\begin{eqnarray}
    \Delta M^2 &=& 2.5\times 10^{-3}~{\rm eV^2} \nonumber
    \\
    \sin^2\theta_{23} &=& 0.49 \, (\sim 0.51) \nonumber
    \\
    \sin^2\theta_{13} &=& 0.005
    \\
    \chi^2_{\rm atm+CHOOZ, min} &=& 40.2 \nonumber
\end{eqnarray}
for 63 d.o.f.
This is the minimum used to obtained the allowed parameters 
in the one-dominant mass scale approximation: the upper panels in 
Fig.~\ref{fig:globalwc} and the ranges in Eq.~\eqref{eq:atmchparam0}.

In the upper panel of Fig.~\ref{fig:chialpha} we show the dependence
of $\Delta\chi^2_{\rm atm+CHOOZ}$ on $\alpha$. From this figure we see
that the inclusion of the CHOOZ reactor data results into a stronger
dependence of the analysis on the value of $\Delta m^2$ (or
equivalently on $\alpha$) and large values of the mass splitting
hierarchy parameter become disfavoured. Also the dependence is
stronger for the Normal scheme, as expected (see discussion below
Eq.~\eqref{eq:choozdif}). As a consequence the ranges of mixing
parameters~-- which, in the analysis of atmospheric data alone, were
broaden in presence of large values of $\alpha$~-- are expected to
become narrower with the inclusion of the CHOOZ data in the analysis.

This effect is explicitly shown in
Fig.~\ref{fig:chiswc}, where we plot the
dependence of the $\Delta\chi^2_{\rm atm+CHOOZ}$ on $\Delta M^2$,
$\theta_{23}$ and $\theta_{13}$, respectively, for different values of
$\alpha$ (to be compared with the corresponding 
Fig.~\ref{fig:chisnc} for the analysis of the
atmospheric data). Fig.~\ref{fig:chiswc} is shown for the Inverted scheme.
(the corresponding figure for the Normal scheme is very similar). 
The figure illustrates that indeed the inclusion
of the CHOOZ data in the analysis results into a reduction of the
allowed ranges for the mixing angles, in particular $\theta_{13}$. 
From this analysis we obtain the $3\sigma$ allowed (1 d.o.f.)
bounds:
\begin{itemize}
\item[$\bullet$] For arbitrary $\alpha$
\begin{equation}
    \begin{tabular}{c@{\hspace{10mm}}c}
        Normal & Inverted \\
        $1.3\leq \frac{\Delta M^2}{10^{-3}~{\rm eV^2}}\leq 5.4$ &
        $1.3\leq \frac{\Delta M^2}{10^{-3}~{\rm eV^2}}\leq 5.2$
        \\
\multicolumn{2}{c} {$\cos\delta=\pm 1$}\\
        $0.26\leq \sin^2\theta_{23}\leq 0.71$ &
        $0.26\leq \sin^2\theta_{23}\leq 0.70$ \\
        $\sin^2\theta_{13}\leq 0.06$ &
        $\sin^2\theta_{13}\leq 0.07$ \\
    \end{tabular}
    \label{eq:atmchparam}
\end{equation}
\item[$\bullet$] For $\alpha=0$
\begin{equation}
\begin{tabular}{c@{\hspace{10mm}}c}
        Normal & Inverted \\
        $1.3\leq \frac{\Delta M^2}{10^{-3}~{\rm eV^2}}\leq 5.1$ &
        $1.3\leq \frac{\Delta M^2}{10^{-3}~{\rm eV^2}}\leq 5.0$ \\
        $0.31\leq \sin^2\theta_{23}\leq 0.71$ &
        $0.31\leq \sin^2\theta_{23}\leq 0.70$ \\
        $\sin^2\theta_{13}\leq 0.07$ &
        $\sin^2\theta_{13}\leq 0.07$\\
    \end{tabular}
    \label{eq:atmchparam0}
\end{equation}
\end{itemize}
Comparing with the results in Eq.~\eqref{eq:atmparam} we see that
including the CHOOZ reactor data reduces the effect on the final
allowed range of parameters arising from allowing departures from the
one mass scale dominance approximation.  In other words the ranges in
Eq.~\eqref{eq:atmchparam} and \eqref{eq:atmchparam0} are not very
different.

Fig.~\ref{fig:globalwc} shows the global 2-dimensional allowed regions
in ($\cos\delta\sin^2\theta_{23}, \Delta M^2$) from the analysis of
the atmospheric neutrino and CHOOZ reactor data for optimized values
$\alpha$ and $\theta_{13}$ as well as the results for the one mass
scale dominance approximation $\alpha=0$ case. Comparison with
Fig.~\ref{fig:globalnc} shows that after including the CHOOZ
reactor data the allowed range of parameters $\Delta M^2$ and
$\sin^2\theta_{23}$ becomes more ``robust'' and it is almost
independent of the Normal or Inverted ordering of the masses or of the
particular choice of $\cos\delta=\pm 1$.

How large would have $\alpha$ and/or $\theta_{13}$ to be for 3--$\nu$ 
effects to be visible in the global analysis?. We find that 
in order to have a 3$\sigma$ effect on the global analysis 
either $\tan^2\theta_{13}$ should be larger than $0.07$ or $\alpha$ 
should be larger than 0.4.

\section{Summary}
\label{sec:sum}

In this article we have explored the effect of keeping the two
mass scales on the three--flavour oscillation analysis of the
atmospheric and reactor neutrino data. 
First we have performed the independent analyses of the atmospheric
neutrino data and of the CHOOZ data. We have studied the allowed
parameter space resulting from these analyses as a function of the
mass splitting hierarchy parameter $\alpha=\Delta m^2/ \Delta M^2$
which parametrizes the departure from the one--dominant mass scale
approximation. Finally we have studied the effect of keeping the two
mass scales on the combined analysis.

In general the analysis of atmospheric data involves six parameters: 
two mass differences, which
we denote by $\Delta M^2$ and $\Delta m^2$, three mixing angles
($\theta_{23}$, $\theta_{13}$ and $\theta_{12}$) and one CP phase
($\delta$). The analysis of the reactor data involves four of these
parameters, namely, $\Delta M^2$ and $\delta m^2$, $\theta_{13}$ and
$\theta_{12}$. For the sake of simplicity we have concentrated on the
dependence on  $\Delta m^2$ while keeping the CP phase fixed to 
CP conserving values and the mixing
angle $\theta_{12}$ to be within the LMA range favoured in the 
global analysis of solar neutrino data by choosing a characteristic 
value $\tan^2\theta_{12}=0.45$. 
We have verified that the atmospheric data alone or in combination 
with the CHOOZ data is not sensitive enough to give any 
constraint on the 
possibility of  CP violation nor to variations of the 
$\tan^2\theta_{12}$ within the allowed LMA range. Thus 
our conclusions are robust.

Our results can be summarized as follows:
\begin{itemize}
  \item The dominant effect of a non-vanishing value of $\alpha$
    in the atmospheric neutrino events is an increase (decrease) of
    the expected number of contained $\nu_e$ for $\theta_{23}$ in the
    first (second octant) as previously discussed in
    Refs.~\cite{smirnov1,smirnov2}.
  \item In the predicted atmospheric neutrino events the effects of a
    non-vanishing $\alpha$ and of the mixing angle $\theta_{13}$ can
    have opposite signs and certain degree of cancellation may occur
    between both effects. 
  \item The survival probability of $\bar\nu_e$ at CHOOZ decreases for
    increasing values of $\theta_{13}$ and $\alpha$, so that the
    effect of both parameters is additive in the CHOOZ reactor data.
    For $\theta_{12}\leq \frac{\pi}{4}$     
    the effect of $\Delta m^2$ is slightly stronger for the Normal
    mass ordering \cite{choozpetcov,irina}.
  \item Allowing for a non-zero value of $\alpha$ very mildly improves
    the quality of the atmospheric neutrino fit as a consequence of
    the better description of the sub-GeV electron data. This effect
    drives the best fit point to the first octant of the mixing angle
    $\theta_{23}$.
  \item  Still the fit to atmospheric neutrinos is very insensitive
    to large values of $\alpha$ as long as all other parameters are
    allowed to vary accordingly.
  \item As a consequence the allowed range of $\sin^2\theta_{13}$ and
    $\sin^2\theta_{23}$ from the atmospheric neutrino data analysis
    becomes, in general, broader than the one for the $\alpha=0$ case.
  \item On the other hand the allowed range of $\Delta M^2$ obtained
    from the atmospheric neutrino data fit is stable under departures
    from the one mass scale dominance approximation.
  \item The inclusion of the CHOOZ reactor data in the analysis
    leads to a stronger dependence of the results on the value of
    $\alpha$, with smaller values of $\alpha$ and $\theta_{13}$ favoured.
  \item As a consequence the final determination of the allowed ranges
    for both $\Delta M^2$ and the mixing angles $\theta_{23}$ and
    $\theta_{13}$ is very robust and the ranges are only slightly
    different from those obtained in the one mass scale dominance
    approximation.
\end{itemize}

\begin{acknowledgement}
    We thank C.\ Pe\~na-Garay and T.\ Schwetz for discussions. The
    work of M.M.\ is supported by the EU contract HPMF-CT-2000-01008.
    MCG-G is supported by the EU contract HPMF-CT-2000-00516. This
    work was also supported by the Spanish DGICYT under grants
    PB98-0693 and FPA2001-3031, by the European Commission RTN network
    HPRN-CT-2000-00148 and by the European Science Foundation network
    grant N.~86.
\end{acknowledgement}

\newpage


\end{document}